# Predicting publication long-term impact through a combination of early citations and journal impact factor

Giovanni Abramo (*corresponding author*)
  *Laboratory for Studies in Research Evaluation*
  *Institute for System Analysis and Computer Science (IASI-CNR)*
  *National Research Council of Italy*
  ADDRESS:  Istituto di Analisi dei Sistemi e Informatica, Consiglio Nazionale delle Ricerche
            Via dei Taurini 19, 00185 Roma - ITALY
            tel. +39 06 7716417, fax +39 06 7716461,
            giovanni.abramo@uniroma2.it

Ciriaco Andrea D'Angelo
  *University of Rome "Tor Vergata" and Institute for System Analysis and Computer Science-National Research Council of Italy*
  ADDRESS:  Dipartimento di Ingegneria dell'Impresa
            Università degli Studi di Roma "Tor Vergata"
            Via del Politecnico 1, 00133 Roma - ITALY
            Tel. and fax +39 06 72597362, dangelo@dii.uniroma2.it
            ORCID: 0000-0002-6977-6611

Giovanni Felici
  *National Research Council of Italy*
  ADDRESS:  Istituto di Analisi dei Sistemi e Informatica, Consiglio Nazionale delle Ricerche
            Via dei Taurini 19, 00185 Roma - ITALY
            tel. +39 06 7716417, fax +39 06 7716461, giovanni.felici@iasi.cnr.it

**Abstract**
The ability to predict the long-term impact of a scientific article soon after its publication is of great value towards accurate assessment of research performance. In this work we test the hypothesis that good predictions of long-term citation counts can be obtained through a combination of a publication's early citations and the impact factor of the hosting journal. The test is performed on a corpus of 123,128 WoS publications authored by Italian scientists, using linear regression models. The average accuracy of the prediction is good for citation time windows above two years, decreases for lowly-cited publications, and varies across disciplines. As expected, the role of the impact factor in the combination becomes negligible after only two years from publication.

# 1. Introduction

Scientific publications encoding new knowledge have different values, depending on their impact on future scientific advancements and ultimately on social and economic development. As a proxy for such impact, bibliometricians adopt citation-based indicators. The underlying assumption is that when a publication is cited, it has contributed to (has had an impact on) the new knowledge encoded in the citing publications (Bornmann & Daniel, 2008). Citational analysis has become the principal instrument for evaluating the bibliometric impact of scientific production, whether used independently or in "informed peer review" evaluation exercises.

The questions for evaluative scientometrics are: i) which indicator (or combination) best predicts future impact; and ii) what is its predictive power. The answers to these queries are seen to vary, depending on the time elapsed from date of publication to measurement of accrued citations. There is general agreement on the fact that late citation counts (as proxy of long-term impact) serve as the benchmark for determining the best indicator or combination of (and its predictive power), for each citation time window (Abramo, 2018).

What becomes clear is that there is an embedded tradeoff between level of accuracy and timeliness in measurement, and also clear is that the scientometrician has the responsibility of communicating this in their relations with decision-makers. Indeed, one of the critical issues in reliability of citation indicators of impact concerns the rapidity with which citations accumulate: citations accrue with time, and no one can know for sure for how much time. Citation count can only serve as a reliable proxy of the scholarly impact of a work if observed at sufficient distance from the date of publication, or applying what is called a "citation time window" of adequate length. Yet, given a performance assessment aimed at informing policy and management decisions, no reasonable decision-maker could wait the necessary decades for completion of the citation life cycle. Hence, in designing a research assessment, the question becomes what length of citation time window should be selected in order for early citations to qualify as an accurate proxy of impact. Inevitably, finding an answer entails addressing the tradeoff between level of accuracy and timeliness in measurement. Previous literature signals that the "best trade-off" would differ across disciplines, since the life cycles of citations and peaks in the citation distribution curves vary with this factor (Garfield, 1972; Mingers, 2008; Wang, 2013; Baumgartner & Leydesdorff, 2014). Adams (2005) states that citations accumulated one and two years after publication "might be useful as a forward indicator of the long-term quality of research publications". Rousseau (1988) and Glänzel, Schlemmer, and Thijs (2003) noted that for mathematics the standard bibliometric time horizon must be greater than for other fields. Abramo, Cicero, and D'Angelo (2011) found that in biology, biomedical research, chemistry, clinical medicine and physics, the peak in citations occurs in the second year after publication; but in earth and space science and in engineering, citations follow a more regular and slower-growing trend. Mathematics behaves still differently, with publications collecting citations very slowly.

Some bibliometricians have investigated the possibility of using alternative metrics, or "altmetrics", to increase accuracy in predicting impact - an area of research now attracting considerable interest. Since reading a publication occurs before citing it, then particularly for the problem of recent publications, it could make sense to count readers (through on-line views, downloads, tweets or other digitally traceable behaviors), rather than relying on citations; it might also be possible to use altmetrics to supplement indicators based on citations (Priem, Taraborelli, Groth, & Neylon, 2010; Li, Thelwall, & Giustini, 2012; Thelwall, Haustein, Larivière, & Sugimoto, 2013; Shema, Bar-Ilan, & Thelwall, 2014; Sud, & Thelwall, 2014; Thelwall & Sud, 2016). In searching to improve the predictive power of early citations, bibliometricians have also proposed combining citation counts with other independent variables. *Science* recently hosted a lively discussion on the topic. Through a mechanistic model collapsing the citation histories of publications from different journals and disciplines into a single curve, Wang, Song and Barabási (2013) concluded that "all papers tend to follow the same universal temporal pattern". Wang, Mei, and Hicks (2014) objected that "their analyses find discouraging results … and correspondingly enormous prediction errors. The prediction power is even worse than simply using short-term citations to approximate long-term citations."

Several years earlier, Abramo, D'Angelo, and Di Costa (2010) had provided evidence that for citation windows of two years or less, the journal's impact factor (IF) is a better predictor of impact than citations, for articles in mathematics (and with weaker evidence in biology and earth sciences). Levitt and Thelwall (2011) verified the predictive power of a combination of journal impact and citations, and recommended the hybrid indicator for citation windows of zero or one year only, with the exception of mathematics (and with weaker evidence in biology and earth sciences) where its use for up to a two-year window has been suggested. Bornmann, Leydesdorff, and Wang (2014) also found that IF can be a significant covariate in predicting the citation impact of individual publications. Stern (2014) confirmed that in the social sciences, IF improves correlation between predicted and actual ranks by citations, when applied in the "zero" year of publication and up to one year afterwards. Stegehuis, Litvak, and Waltman (2015) proposed a model to predict a probability distribution for the future number of citations of a publication, using the IF of the hosting journal and the number of citations received by the publication within one year of appearance. In the latest Italian research assessment exercise (VQR 2011-2014) the choice was made to adopt a linear weighted combination of citations and journal metric percentiles, with weights differentiated by discipline and year; but such combination appears to have been formulated without rigorous scientific method. Abramo and D'Angelo (2016) demonstrated that the proposed weighting provides a worse prediction of the impact of publications than the simple citation count.

In this work, we try to identify the combination of IF and early citation counts that best predicts long-term citations of publications in each discipline, by using two alternative prediction models, one based on the rescaled citation counts and another utilizing the log-transformed citation counts. We also analyze the error distribution for the prediction as a function of the number of early citations. Differently from previous contributions in the literature, we also provide: i) the weighted combinations of citations and IF, as a function of the citation time window and field of research, which best predict future impact; and ii) the predictive power of each combination. These two main contributions of the current work feed into the ultimate aim: providing scholars and practitioners with a tool supporting effective design and implementation of research assessments.

## 2. Data and methods

Our reference framework is the Italian national research assessment exercise, intended to assess the performance of research institutions through the evaluation of their research products from a period of time. Bibliometric evaluation can be applied to research products indexed in such repositories as Web of Science (WoS) and Scopus. The publications under evaluation are issued in different years, and belong to different fields or subject categories (SCs). We approximate the long-term impact of an article by the number of citations counted nine years after publication. For example, for papers published in 2004 (2005, 2006), we measure the long term impact with the number of citations received up to year 2013 (2014, 2015).[1] We then regress the long-term impact on the number of citations accrued in previous years and on the IF of the journal at publication date.[2] In doing so, we adopt two different linear models.

In the first model, we rescale the number of citations by year and subject category, an approach frequently used by bibliometricians when comparing publications of different years and fields. In line with Abramo, Cicero, and D'Angelo (2012), we rescale citations (respectively IFs) to the average of all cited Italian publications (respectively journals) indexed by WoS in the same year and SC. Then we regress the long-term impact on the rescaled number of citations accrued in previous years and on the rescaled IF of the journal at publication date. More formally, we define the rescaled variables as follows:

$$y_t^i = \frac{c_t^i}{\bar{c}_t}$$

$$x^i = \frac{IF_k^i}{\overline{IF}_k}$$

where:
- $t$ is the citation time window, with $0 \leq t \leq 8$;
- $c_t^i$ is the number of citations received by publication $i$, $t$ years after publication;
- $\bar{c}_t$ is the average number of citations received $t$ years after publication by all cited publications of the same year and SC of publication $i$;
- $k$ is the publication year, with $2004 \leq k \leq 2006$;
- $IF_k^i$ is the impact factor of the journal hosting publication $i$, at publication year;
- $\overline{IF}_k$ is the average of impact factors of all journals falling in the SC of publication $i$, at publication year.

We adopt a linear regression model of the following form:

$$y_{k+9}^i \approx b_0^t + b_1^t x^i + b_2^{k+t} y_{k+t}^i$$

[1]

The second model follows an alternative approach based on the logarithmic transformation of the citation counts. The underlying reason is that the distribution of citations to articles in a given year and SC is likely to be approximately log-normally distributed (Stringer, Sales-Pardo, & Amaral, 2008). Such approach has the effect of reducing the magnitude of the number of citations in a non-proportional way, diminishing the size of the large citation counts and controlling for their variance, with potentially beneficial effects for the application of a linear regression model. Of course, such transformation has an effect also on the interpretation of the regression coefficient, that would approximate the linear relation between the percentage increases of the dependent and the independent variables. In this case we have:

$$y_t^i = \log(1 + c_t^i)$$

---
[1] We are able to observe citations accrued by 31 December of each year, until 2015.
[2] The two-year IF.

[2]

and the same regression equation represented in [1].

In both models, the obvious autocorrelation of the number of citations makes the role of $y_{k+t}^i$ more and more important in explaining $y_{k+9}^i$ as $t$ approaches $8$. Additionally, the model based on rescaled citations may as well be interpreted in terms of the explanation of the relative increase of citations of the paper between year $t$ and year 9. For the reason above we expect that both models proposed would have a very good fit for large values of $t$, exploiting the autocorrelation between the dependent and one of the two independent variables. Nevertheless, we believe it is interesting to find out when this autocorrelation effect starts to be strong enough to achieve accurate prediction of long term impact, and how it varies as a function of the SCs and the size of early citations. Last but not least, we are interested in assessing the relative weight of early citations and IF, as $t$ varies.

To calibrate the regression models and answer our research questions, we consider a dataset consisting of all Italian publications indexed in WoS over 2004-2006 in the sciences and social sciences (only articles, reviews, conference proceedings and letters, totaling 123,128 items). We exclude the Art & Humanities SCs because of the limited coverage of outputs by WoS. We assign each publication to the SC of the hosting journal, according to the WoS classification scheme. Publications in multi-category journals are assigned to each of the SCs. We run regression models for $t = 0, \ldots, 8$ for each of the 170 SCs with more than 100 publications, using the *lm* function in package MASS of R version 3.2.2.

In some cases, the regressions models for specific subgroups resulted positive to heteroskedasticity tests (such as the Breusch-Pagan test). All regression coefficients presented in the following were thus obtained by a robust standard error approach for the computation of the p-values (heteroskedasticity consistent variance-covariance matrix HC3, see Zeileis (2004) for details).

We tested also more complex quadratic regressions and non-parametric classification models (decision trees) but without significant improvement of fitting with respect to the linear regression shown in [1].

## 3. Results

To start, as an example, we apply the OLS regressions to the 1113 Italian publications falling in the SC "Engineering, chemical" (Table 1). Each row shows results for each citation time window. All early citations coefficients exhibit a high statistical significance, as attested by the associated *p-values*, while the IF coefficients are not always significant. $R^2$ increases steadily with the citation time window for both models. Specifically, the values of the coefficient associated to the IF (3[rd] and 7[th] columns of the Table 1) decrease drastically in both cases, becoming very small and negative respectively after three and five years from publication.

Differently, early citations coefficients are large and steadily increasing in the first three years after publication, further weakening the weight of the IF in predicting long-term citations. We can conclude that with a citation time window of three years the role of the IF becomes negligible in both models, which exhibit very good fit ($R^2$=0.812 and $R^2$=0.790 respectively).

Regression results for each of the 170 SCs can be found in Appendix A, for a three-year citation window (or in Supplementary Material-SM_1, for all citation time windows).[3] They are summarized in Table 2 and Table 3 where, for the reference three-year citation window, we show descriptive statistics of results for the SCs of each macro-area, for the two OLS models. Regression coefficients for the IF show a *p-value* lower than 0.1 in 40 SCs (out of 170) in the OLS on rescaled citation counts (third column of Table 2) and in 102, in the OLS on log-transformed citations (third column of Table 3). The maximum value of the IF coefficient is in Engineering (0.50 for the rescaled citations, 0.31 for the log-transformed ones), while the average values are small in size and negative in seven macro-areas in the model with rescaled citations. On the other hand, the coefficients of early citations are always significant at 0.01 level for both models. $R^2$ averages 0.8, with lowest average value for Mathematics (0.692 for rescaled citations, 0.717 for log-transformed ones), and highest for Multidisciplinary Sciences (0.831 and 0.828) and Clinical Medicine (0.828 and 0.847). Apart from minimal divergence in the rankings, the two models provide very similar outcomes.

*Table 1: Regressions for 1113 publications in "Engineering, chemical" and citation time window from 0 to 8 years*

| Time window (years) | OLS regression on rescaled citations | | | | OLS regression on log-transformed citations | | | |
|---|---|---|---|---|---|---|---|---|
| | Intercept | Impact Factor coeff. | Early citations coeff. | $R^2$ | Intercept | Impact Factor coeff. | Early citations coeff. | $R^2$ |
| 0 | 0.087 | 0.475*** | 0.661*** | 0.157 | 0.757*** | 0.19*** | 0.584*** | 0.247 |
| 1 | 0.107 | 0.174*** | 0.968*** | 0.447 | 0.671*** | 0.104*** | 0.876*** | 0.462 |
| 2 | 0.065 | 0.024 | 1.088*** | 0.672 | 0.517*** | 0.044*** | 0.985*** | 0.662 |
| 3 | 0.051 | -0.040 | 1.127*** | 0.812 | 0.374*** | 0.016** | 1.035*** | 0.790 |
| 4 | 0.023 | -0.047 | 1.121*** | 0.886 | 0.255*** | 0.004 | 1.055*** | 0.877 |
| 5 | 0.026 | -0.053** | 1.104*** | 0.930 | 0.18*** | -0.002 | 1.047*** | 0.922 |
| 6 | 0.014 | -0.038** | 1.080*** | 0.961 | 0.118*** | -0.004 | 1.039*** | 0.951 |
| 7 | 0.008 | -0.024** | 1.051*** | 0.983 | 0.074*** | -0.004* | 1.023*** | 0.975 |
| 8 | 0.005 | -0.013*** | 1.026*** | 0.994 | 0.031*** | -0.003** | 1.013*** | 0.990 |

*Dependent variable: rescaled citations nine years after publication (column 2-5 model); log-transformed citations nine years after publication (column 6-9 model).*
*Statistical significance: \*p-value <0.1, \*\*p-value <0.05, \*\*\*p-value <0.01.*

---

[3] We focus on the three-year citation window because it is the minimum citation time window in the Italian research assessment exercise, and is also the one whereby the model fit starts being acceptable.

*Table 2: Regressions statistics (OLS on rescaled citations) for subject categories in each macro-area, for a three-year citation window*

| Macro-area | SCs | With IF p-value < 0.1 | Impact Factor coefficients† | | | Early citations coefficients | | | $R^2$ | |
|---|---|---|---|---|---|---|---|---|---|---|
| | | | Min;Max | Mean | St. Dev. | Min;Max | Mean | St. Dev. | Mean | St. Dev. |
| Biology | 27 | 9 | [-0.15;0.22] | 0.04 | 0.14 | [0.83;1.46] | 1.08 | 0.16 | 0.81 | 0.05 |
| Biomedical research | 14 | 1 | [0.06;0.06] | 0.06 | n.a. | [0.74;1.6] | 1.07 | 0.22 | 0.81 | 0.10 |
| Chemistry | 8 | 3 | [-0.05;0.07] | -0.01 | 0.06 | [0.99;1.50] | 1.16 | 0.19 | 0.81 | 0.06 |
| Clinical medicine | 35 | 4 | [-0.16;0.21] | 0.01 | 0.18 | [0.91;1.53] | 1.13 | 0.16 | 0.83 | 0.10 |
| Earth and space sciences | 12 | 2 | [0.10;0.13] | 0.11 | 0.02 | [0.87;1.23] | 1.06 | 0.11 | 0.81 | 0.06 |
| Economics | 4 | 1 | [0.22;0.22] | 0.22 | n.a. | [0.89;1.27] | 1.08 | 0.17 | 0.73 | 0.03 |
| Engineering | 35 | 12 | [-0.30;0.50] | 0.06 | 0.22 | [0.88;1.69] | 1.18 | 0.22 | 0.77 | 0.06 |
| Law, political and social sc. | 4 | 0 | - | - | - | [1.02;1.40] | 1.16 | 0.17 | 0.80 | 0.05 |
| Mathematics | 6 | 2 | [0.15;0.16] | 0.15 | 0.00 | [0.79;1.42] | 1.11 | 0.20 | 0.69 | 0.05 |
| Multidisciplinary sciences | 2 | 1 | [-0.02;-0.02] | -0.02 | n.a. | [0.95;1.23] | 1.09 | 0.20 | 0.83 | 0.18 |
| Physics | 18 | 4 | [-0.31;0.22] | -0.07 | 0.26 | [0.71;1.62] | 1.13 | 0.27 | 0.82 | 0.09 |
| Psychology | 5 | 1 | [-0.11;-0.11] | -0.11 | n.a. | [1.06;1.24] | 1.15 | 0.08 | 0.82 | 0.05 |

*Dependent variable: value of rescaled citation counts nine years after publication*
*† Statistics computed considering only SCs with IF coefficient p-values lower than 0.1*

*Table 3: Regressions statistics (OLS on log-transformed citations) for subject categories in each macro-area, for a three-year citation window*

| Macro-area | SCs | With IF p-value < 0.1 | Impact Factor coefficients† | | | Early citations coefficients | | | $R^2$ | |
|---|---|---|---|---|---|---|---|---|---|---|
| | | | Min;Max | Mean | St. Dev. | Min;Max | Mean | St. Dev. | Mean | St. Dev. |
| Biology | 27 | 19 | [-0.04;0.10] | 0.02 | 0.04 | [0.90;1.08] | 0.99 | 0.05 | 0.80 | 0.04 |
| Biomedical research | 14 | 5 | [-0.01;0.01] | 0.00 | 0.01 | [0.98;1.10] | 1.05 | 0.04 | 0.85 | 0.05 |
| Chemistry | 8 | 7 | [0.01;0.03] | 0.02 | 0.01 | [0.96;1.07] | 1.01 | 0.03 | 0.81 | 0.04 |
| Clinical medicine | 35 | 15 | [-0.01;0.07] | 0.01 | 0.02 | [0.98;1.21] | 1.08 | 0.06 | 0.85 | 0.05 |
| Earth and space sciences | 12 | 7 | [0.01;0.07] | 0.04 | 0.03 | [0.84;1.21] | 1.04 | 0.09 | 0.78 | 0.04 |
| Economics | 4 | 0 | - | - | - | [1.01;1.29] | 1.17 | 0.11 | 0.76 | 0.04 |
| Engineering | 35 | 26 | [-0.08;0.31] | 0.07 | 0.08 | [0.91;1.20] | 1.06 | 0.06 | 0.77 | 0.06 |
| Law, political and social sc. | 4 | 2 | [-0.04;0.07] | 0.01 | 0.08 | [1.16;1.41] | 1.23 | 0.12 | 0.81 | 0.07 |
| Mathematics | 6 | 5 | [0.03;0.17] | 0.07 | 0.06 | [0.94;1.10] | 1.06 | 0.06 | 0.72 | 0.06 |
| Multidisciplinary sciences | 2 | 1 | [0.00;0.00] | 0.00 | n.a. | [1.20;1.26] | 1.23 | 0.04 | 0.83 | 0.21 |
| Physics | 18 | 13 | [-0.06;0.08] | 0.01 | 0.03 | [1.01;1.14] | 1.06 | 0.03 | 0.82 | 0.06 |
| Psychology | 5 | 2 | [-0.03;0.01] | -0.01 | 0.03 | [0.94;1.27] | 1.07 | 0.12 | 0.80 | 0.04 |

*Dependent variable: value of log-transformed citation counts nine years after publication (second model)*
*† Statistics computed considering only SCs with IF coefficient p-values lower than 0.1*

Table 4 and Table 5 provide the full details of the regressions for the 10 SCs with the highest and lowest $R^2$, for a three-year citation window (complete results are available in Appendix A), in the two models.

As for rescaled citations (Table 4) only "Neuroimaging" e "Radiology, Nuclear Medicine & Medical Imaging" have $R^2$ values below 0.6; 94 SCs have $R^2$ values above 0.8, with six of the top listed subject categories belonging to Clinical medicine. Also for log-trasformed citations (Table 5) six of the top listed subject categories belong to Clinical medicine, while six at the bottom belong to Engineering.

As expected, the two models may perform differently at individual SC level; no specific patterns of performance emerge at a preliminary analysis.

In order to better appreciate differences across macro-areas, Figure 1 shows the dispersion of average regression coefficients for SCs in each macro-area, for a three-year citation window for the two models. Analysing the two panels some considerations may be drawn:

- Economics shows (on average) the maximum relative weight of IF with respect to early citations in both models;
- Psychology is on the opposite side in both models;
- Life science areas (biomedical research, chemistry, biology, clinical medicine) have varying average early citation coefficients but very similar IF average coefficients, typically very small;
- Law and political sciences seem to experience a strong effect of early citations in both models; similarly, Engineering (in the left panel) and Multidisciplinary sciences (in right panel).

*Table 4: Regression results (OLS on rescaled citations) for 10 best and the 10 worst SCs according to regression fit in model ($R^2$, last column of the table), for a three-year citation window*

| Subject Category | Macro Area | Obs. | Intercept | Impact Factor coeff. | Early citations coeff. | $R^2$ |
|---|---|---|---|---|---|---|
| Physics, nuclear | Physics | 1548 | 0.024 | 0.229 | 0.706** | 0.972 |
| Medicine, general & internal | Clinical medicine | 418 | -0.031 | 0.004 | 1.124*** | 0.964 |
| Multidisciplinary sciences | Multidisciplinary sciences | 126 | -0.034 | -0.016* | 1.233*** | 0.958 |
| Peripheral vascular disease | Clinical medicine | 1959 | -0.064*** | 0.001 | 1.110*** | 0.925 |
| Physics, particles & fields | Physics | 3690 | 0.033 | 0.15 | 0.755** | 0.923 |
| Nutrition & dietetics | Clinical medicine | 853 | -0.109 | -0.053 | 1.271*** | 0.918 |
| Genetics & heredity | Clinical medicine | 2253 | -0.005 | 0.043 | 0.962*** | 0.917 |
| Endocrinology & metabolism | Clinical medicine | 2985 | 0.013 | 0.036 | 0.979*** | 0.913 |
| Meteorology & atmospheric sciences | Earth and space sciences | 868 | -0.109** | 0.046 | 1.108*** | 0.910 |
| Urology & nephrology | Clinical medicine | 1713 | 0.012 | -0.009 | 1.084*** | 0.909 |
| ... | | | | | | |
| Engineering, ocean | Engineering | 109 | -0.04 | 0.099** | 0.984*** | 0.684 |
| Engineering, industrial | Engineering | 299 | 0.035 | 0.052 | 1.144*** | 0.681 |
| Mineralogy | Earth and space sciences | 406 | 0.059 | 0.097* | 0.867*** | 0.680 |
| Logic | Mathematics | 113 | 0.152* | 0.067 | 0.790*** | 0.679 |
| Materials science, multidisciplinary | Engineering | 3568 | -0.298 | -0.112 | 1.689*** | 0.665 |
| Physics, condensed matter | Physics | 3565 | -0.38 | -0.026 | 1.553*** | 0.664 |
| Physics, applied | Physics | 3879 | -0.324 | -0.071 | 1.558*** | 0.650 |
| Statistics & probability | Mathematics | 686 | 0.03 | -0.148 | 1.415*** | 0.612 |
| Radiology, nuclear medicine & med. imaging | Biomedical research | 1856 | -0.043 | 0.062* | 1.024*** | 0.514 |
| Neuroimaging | Clinical medicine | 308 | -0.112 | -0.076 | 1.517*** | 0.325 |

*Dependent variable: value of rescaled citation counts nine years after publication.*
*Statistical significance: \*p-value <0.1, \*\*p-value <0.05, \*\*\*p-value <0.01.*

*Table 5: Regression results (OLS on log-transformed citations) for 10 best and the 10 worst SCs according to regression fit in model ($R^2$, last column of the table), for a three-year citation window*

| Subject Category | Macro Area | Obs. | Intercept | Impact Factor coeff. | Early citations coeff. | $R^2$ |
|---|---|---|---|---|---|---|
| Multidisciplinary sciences | Multidisciplinary sciences | 126 | 0.194*** | -0.004*** | 1.198*** | 0.974 |
| Medicine, general & internal | Clinical medicine | 418 | 0.108*** | -0.001** | 1.182 | 0.962 |
| Neuroimaging | Clinical medicine | 308 | 0.130*** | 0.044*** | 1.119*** | 0.916 |
| Allergy | Biomedical research | 429 | 0.256*** | -0.002 | 1.095*** | 0.907 |
| Astronomy & astrophysics | Physics | 4694 | 0.214*** | 0.002*** | 1.060*** | 0.906 |
| Peripheral vascular disease | Clinical medicine | 1959 | 0.240*** | 0.000 | 1.113*** | 0.904 |
| Gastroenterology & hepatology | Clinical medicine | 2102 | 0.253*** | -0.007*** | 1.137*** | 0.903 |
| Cardiac & cardiovascular systems | Clinical medicine | 2986 | 0.241*** | -0.003*** | 1.133*** | 0.902 |
| Hematology | Biomedical research | 2704 | 0.275*** | -0.001 | 1.074*** | 0.900 |
| Critical care medicine | Clinical medicine | 519 | 0.264*** | -0.012*** | 1.157*** | 0.899 |
| ... | | | | | | |
| Thermodynamics | Physics | 504 | 0.417*** | 0.000 | 1.055*** | 0.690 |
| Engineering, civil | Engineering | 725 | 0.453*** | 0.057*** | 1.000*** | 0.688 |
| History & philosophy of science | Multidisciplinary sciences | 115 | 0.039 | 0.190 | 1.259*** | 0.682 |
| Mineralogy | Earth and space sciences | 406 | 0.468*** | 0.054*** | 0.842*** | 0.679 |
| Engineering, aerospace | Engineering | 415 | 0.288*** | -0.037** | 1.083*** | 0.671 |
| Engineering, industrial | Engineering | 299 | 0.461*** | 0.075 | 0.979*** | 0.660 |
| Construction & building technology | Engineering | 220 | 0.415*** | 0.226*** | 0.991*** | 0.655 |
| Engineering, manufacturing | Engineering | 319 | 0.529*** | 0.025 | 0.918*** | 0.645 |
| Engineering, geological | Engineering | 230 | 0.424*** | 0.172** | 0.906*** | 0.639 |
| Logic | Mathematics | 113 | 0.249*** | 0.071* | 0.936*** | 0.605 |

*Dependent variable: value of log-transformed citation counts nine years after publication.*
*Statistical significance: *p-value <0.1, **p-value <0.05, ***p-value <0.01.*

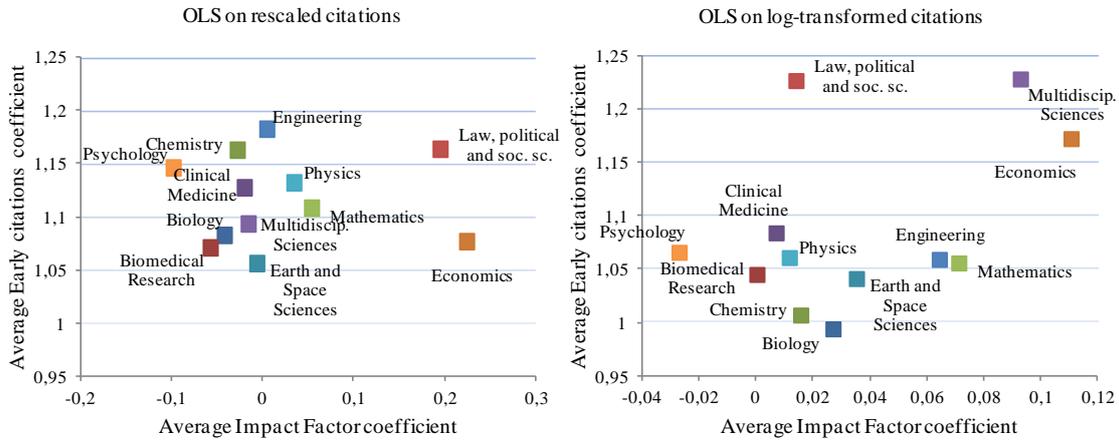

*Figure 1: Dispersion of macro-areas according to their average early citation coefficient and average impact factor coefficient.*
*Dependent variable: Left panel, rescaled citations nine years after publication.*
*Right panel, log-transformed citations nine years after publication.*

Table 6 provides the full details of the regressions for publications that have received no citations within the considered time window. In this case, the impact prediction is based on IF coefficient only. The fitting, as attested by negligible $R^2$ values, is very poor; the IF coefficients markedly decrease in size with the time windows for both models. We run additional regressions for uncited publications also at SC level, for all SCs with more than 50 observations (119 in all), for the three-year citation window only. Results are shown in detail in Appendix B.[4] Only 62 SCs show significant IF coefficients, of which 52 are positive, but their $R^2$ is above 0.2 in only 4 SCs.

*Table 6: Regression results for uncited publications and citation time window from 0 to 8 years*

| Time window (years) | Obs. | OLS regression on rescaled citations | | OLS regression on log-transformed citations | |
|---|---|---|---|---|---|
| | | Impact Factor coeff. | $R^2$ | Impact Factor coeff. | $R^2$ |
| 0 | 94428 | 0.211*** | 0.047 | 0.042*** | 0.054 |
| 1 | 43770 | 0.023*** | 0.003 | 0.002* | 0.000 |
| 2 | 24467 | 0.000 | 0.000 | -0.006*** | 0.003 |
| 3 | 17162 | -0.003*** | 0.001 | -0.006*** | 0.005 |
| 4 | 13618 | -0.003*** | 0.002 | -0.004*** | 0.004 |
| 5 | 11601 | -0.002*** | 0.003 | -0.003*** | 0.004 |
| 6 | 10294 | -0.001*** | 0.002 | -0.002*** | 0.002 |
| 7 | 9361 | -0.001*** | 0.002 | -0.001*** | 0.002 |
| 8 | 8640 | 0.000*** | 0.001 | 0.000*** | 0.001 |

*Dependent variable: rescaled citations nine years after publication (column 3-4 model); log-transformed citations nine years after publication (column 5-6 model).*
*Statistical significance: *p-value <0.1, **p-value <0.05, ***p-value <0.01.*

The accuracy of impact prediction depends anyway on the size of early citations. To further investigate this aspect, for each citation time window, we assign each publication to the relevant quartile of the distribution of cited publications (leaving aside the subset of publications that have received no citations at the citation time window *t*). We then run the regressions for each quartile. Results (Table 7) show that the fitting is satisfactory only for Q4 publications. Q2 publications, whose $R^2$ is the lowest, are the most problematic in terms of prediction. Results of all the analyses conducted for each single SC are shown in SM_2, for both models.

*Table 7: Regression results for quartiles of citedness of publications and time window from 0 to 8 years (publications not yet cited within the relevant time window are excluded).*

| Set | Time window (years) | OLS regression on rescaled citations | | | | OLS regression on log-transformed citations | | | |
|---|---|---|---|---|---|---|---|---|---|
| | | Obs. | Impact Factor coeff. | Early citations coeff. | $R^2$ | Obs. | Impact Factor coeff. | Early citations coeff. | $R^2$ |
| Q1 | 0 | 7245 | 0.329*** | 2.780*** | 0.120 | 16810 | 0.046*** | 0.000*** | 0.119 |
| Q2 | 0 | 7163 | 0.457*** | 0.896* | 0.069 | 7163 | 0.046*** | 0.000*** | 0.000 |
| Q3 | 0 | 7169 | 0.394*** | 0.362** | 0.067 | 5794 | 0.038*** | 0.000*** | 0.151 |
| Q4 | 0 | 7184 | 0.463*** | 1.595*** | 0.361 | 6157 | 0.022*** | 0.921*** | 0.370 |
| ALL | 0 | 123189 | 0.280*** | 1.233*** | 0.279 | 123189 | 0.038*** | 0.945*** | 0.234 |
| Q1 | 1 | 19979 | 0.045*** | 1.934*** | 0.063 | 25056 | 0.014*** | 0.000*** | 0.008 |
| Q2 | 1 | 19758 | 0.111*** | 0.650*** | 0.024 | 15458 | 0.023*** | 0.000*** | 0.023 |
| Q3 | 1 | 19843 | 0.124*** | 0.845*** | 0.049 | 22084 | 0.017*** | 0.932*** | 0.083 |
| Q4 | 1 | 19839 | 0.207*** | 1.070*** | 0.532 | 16821 | 0.009*** | 0.987*** | 0.494 |
| ALL | 1 | 123189 | 0.102*** | 1.049*** | 0.581 | 123189 | 0.008*** | 1.055*** | 0.552 |
| Q1 | 2 | 24716 | 0.011*** | 1.381*** | 0.092 | 32508 | 0.000 | 1.073*** | 0.093 |
| Q2 | 2 | 24662 | 0.043*** | 0.953*** | 0.055 | 20598 | 0.005*** | 0.987*** | 0.040 |
| Q3 | 2 | 24698 | 0.043*** | 0.925*** | 0.081 | 21603 | 0.004*** | 0.971*** | 0.088 |
| Q4 | 2 | 24646 | 0.136* | 0.983*** | 0.638 | 24013 | 0.003*** | 1.045*** | 0.645 |

---

[4] For uncited publications, OLS regressions applied to log-transformed citations fail in a number of SCs due to the limited variance of relevant distributions.

| | | | | | | | | |
|---|---|---|---|---|---|---|---|---|
| ALL | 2 | 123189 | 0.064* | 0.999*** | 0.705 | 123189 | -0.001** | 1.081*** | 0.744 |
| Q1 | 3 | 26557 | 0.004*** | 1.095*** | 0.133 | 35396 | -0.003*** | 1.078*** | 0.239 |
| Q2 | 3 | 26477 | 0.018*** | 0.959*** | 0.099 | 22134 | -0.001 | 1.076*** | 0.105 |
| Q3 | 3 | 26491 | 0.015*** | 0.979*** | 0.153 | 23644 | -0.002*** | 1.028*** | 0.168 |
| Q4 | 3 | 26502 | 0.069 | 1.035*** | 0.742 | 24853 | 0.000 | 1.057*** | 0.759 |
| ALL | 3 | 123189 | 0.029 | 1.037*** | 0.797 | 123189 | -0.003*** | 1.077*** | 0.846 |
| Q1 | 4 | 27524 | 0.001** | 1.003*** | 0.216 | 29332 | -0.003*** | 1.076*** | 0.308 |
| Q2 | 4 | 27263 | 0.010*** | 0.945*** | 0.172 | 26076 | -0.002*** | 1.063*** | 0.214 |
| Q3 | 4 | 27406 | 0.005*** | 0.977*** | 0.237 | 27070 | -0.004*** | 1.049*** | 0.301 |
| Q4 | 4 | 27378 | 0.017 | 1.085*** | 0.830 | 27093 | -0.001*** | 1.056*** | 0.840 |
| ALL | 4 | 123189 | 0.004 | 1.069*** | 0.869 | 123189 | -0.003*** | 1.065*** | 0.904 |
| Q1 | 5 | 28055 | 0.000 | 0.984*** | 0.344 | 32148 | -0.002*** | 1.056*** | 0.499 |
| Q2 | 5 | 27756 | 0.003*** | 0.965*** | 0.268 | 26181 | -0.003*** | 1.048*** | 0.320 |
| Q3 | 5 | 27884 | 0.003** | 0.977*** | 0.358 | 26808 | -0.005*** | 1.052*** | 0.431 |
| Q4 | 5 | 27893 | -0.014 | 1.106*** | 0.896 | 26451 | -0.001*** | 1.050*** | 0.900 |
| ALL | 5 | 123189 | -0.011 | 1.082*** | 0.920 | 123189 | -0.003*** | 1.051*** | 0.941 |
| Q1 | 6 | 28241 | 0.000 | 0.971*** | 0.482 | 28730 | -0.002*** | 1.043*** | 0.605 |
| Q2 | 6 | 28241 | 0.000 | 0.977*** | 0.408 | 28417 | -0.003*** | 1.034*** | 0.500 |
| Q3 | 6 | 28189 | -0.001 | 0.989*** | 0.513 | 28599 | -0.004*** | 1.028*** | 0.592 |
| Q4 | 6 | 28224 | -0.024 | 1.099*** | 0.944 | 27149 | -0.001*** | 1.039*** | 0.944 |
| ALL | 6 | 123189 | -0.016 | 1.075*** | 0.957 | 123189 | -0.002*** | 1.037*** | 0.966 |
| Q1 | 7 | 28538 | 0.000 | 0.982*** | 0.666 | 31546 | -0.001*** | 1.032*** | 0.789 |
| Q2 | 7 | 28387 | -0.001 | 0.980*** | 0.589 | 28165 | -0.003*** | 1.025*** | 0.665 |
| Q3 | 7 | 28463 | -0.002*** | 0.992*** | 0.690 | 26647 | -0.003*** | 1.020*** | 0.723 |
| Q4 | 7 | 28440 | -0.024* | 1.074*** | 0.977 | 27470 | -0.001*** | 1.027*** | 0.974 |
| ALL | 7 | 123189 | -0.015** | 1.057*** | 0.982 | 123189 | -0.001*** | 1.024*** | 0.982 |
| Q1 | 8 | 28648 | 0.000 | 0.988*** | 0.854 | 29201 | -0.001*** | 1.014*** | 0.903 |
| Q2 | 8 | 28651 | 0.000 | 0.987*** | 0.808 | 30077 | -0.001*** | 1.010*** | 0.860 |
| Q3 | 8 | 28622 | -0.001*** | 0.996*** | 0.869 | 26661 | -0.001*** | 1.010*** | 0.879 |
| Q4 | 8 | 28628 | -0.014** | 1.039*** | 0.995 | 28610 | -0.001*** | 1.013*** | 0.992 |
| ALL | 8 | 123189 | -0.008*** | 1.030*** | 0.995 | 123189 | -0.001*** | 1.012*** | 0.994 |

*Dependent variable: rescaled citations nine years after publication (column 3-6 model); log-transformed citations nine years after publication (column 7-10 model).*
*Statistical significance: \*p-value <0.1, \*\*p-value <0.05, \*\*\*p-value <0.01.*

We have deepened the analysis through an even finer classification of citedness, assigning each publication to the relevant percentile of the distribution of cited publications. Considering the impact prediction error $E$ for cited-publication $i$ at time window $t$:

$$E_t^i = \left| \frac{y_{k+9}^i - (b_0^t + b_1^t x^i + b_2^{k+t} y_{k+t}^i)}{y_{k+9}^i} \right|$$

[3]

we computed the median error and plotted it in Figure 2. For the model based on rescaled citations we adopted a fine-grained stratification based on deciles (left panel), while for the log-transformed citations we had to resort to quintiles (right panel) as a finer-grained stratification cannot be computed (as a matter of fact, the log-transformed citations quintiles cannot be computed for the case $t = 0$ and $t = 1$ for lack of variability in the citation counts, and are replaced by the overall median error over all quintiles).

As expected, the median of E is higher for publications in the initial percentiles by citations, and then decreases for the subsequent ones, for citation time windows of one year and more (for rescaled citations) and of two years and more (for log-transformed citations). For the three-year citation window the overall median error is 0.40 for rescaled citations, ranging from 0.237 for most cited publications to 0.515 for the least cited (purple line in the left panel). Similarly, we have an overall median error of 0.340, with 0.174 for the most cited publications and 0.407 for the least cited, in the model based on log-transformed citations (purple line in the right panel). Improvements in error may be obtained by limiting the regressions only to cited publications, but such improvements are indeed negligible.

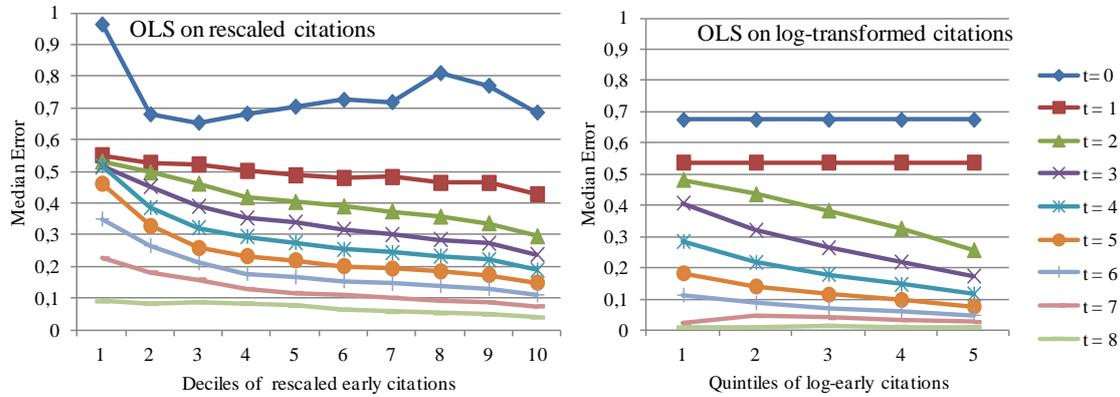

*Figure 2: Median impact prediction error for different citation time windows (t) and different percentiles of citation distribution.*
*Dependent variable: Left panel, rescaled citations nine years after publication (decile 1 = 10% least cited publications; decile 10 = 10% most cited); Right panel, log-transformed citations nine years after publication (quintile 1 = 20% least cited publications; quintile 5 = 20% most cited).*

## 4. Conclusions

The analysis provides several interesting indications for the early assessment of impact of scientific output.

First, we provide statistical evidence that, on average, the choice of a three-year citation window is sufficient to predict the long-term impact of scientific publications with acceptable accuracy, using a linear regression model. A complementary conclusion is that the IF has a non-negligible role only with very short time windows (0 to 2 years); for longer ones, the weight of early citations is dominating and the IF is not informative in explaining the difference between long-term and short-term citations. Second, statistical evidence is also present indicating that the long-term impact of publications with low early citations cannot be predicted with the same accuracy as for those with high early citations. This demands that special attention be taken when papers with a null or small number of early citations need to be evaluated, particularly considering their substantial shares: in our dataset, they are 76.7% for a zero-year citation window, around 13.9% for the reference three-year window, down to 7% for a nine-year window. The decision maker faces an uneasy choice: to apply bibliometrics for evaluation of all publications, which entails accepting high error rates for a noticeable share, or to recur to peer-review for uncited ones, with consequences of higher costs and longer execution times. To any extent, the IF does not help in any way in assessing the impact of such publications. A probably better choice is only to analyse the papers with a threshold level of citations and note the fraction that fall below the threshold. Some papers with good long-term impact will be omitted, but assuming that we are analysing a discipline at the university level there is not a reason to think that the fraction of low citation papers that eventually get high citations should vary systematically across universities.

These idiosyncracies of individual papers should be averaged away at the department/university level (Bruns & Stern, 2016).

Third, both the relative weight of regressors and the accuracy of prediction vary greatly across macro-areas, and across SCs within the macro-areas.

In the future, alternative metrics or "altmetrics" (on-line views, downloads, tweets, other digitally traceable behaviors) could serve as covariates, to improve predictive power when the time window is too short and early citations are too few (Shema, Bar-Ilan, & Thelwall, 2014). Unfortunately, at the time of preparing this study, recent records of altmetrics data were still insufficient to serve such purposes.

Our work draws inspiration from real practical issues, faced by all practitioners called to design a national research assessment exercise with a short citation time window. Given any such national framework, it is appropriate to use the relevant national distributions to assess the impact of research products, as we have done in the current example. However, given its utmost simplicity, the approach proposed in this paper can be replicated in various other contexts, accounting for the specific objectives of the individual assessment exercise.

**References**

Abramo, G. (2018). Revisiting the scientometric conceptualization of impact and its measurement. *Journal of Informetrics,* 12(3), 590-597.

Abramo, G., Cicero, T., & D'Angelo, C.A. (2011). Assessing the varying level of impact measurement accuracy as a function of the citation window length. *Journal of Informetrics,* 5(4), 659-667.

Abramo, G., Cicero, T., D'Angelo, C.A. (2012). Revisiting the scaling of citations for research assessment. *Journal of Informetrics*, 6(4), 470-479.

Abramo, G., D'Angelo, C.A. (2016). Refrain from adopting the combination of citation and journal metrics to grade publications, as used in the Italian national research assessment exercise (VQR 2011-2014). *Scientometrics*, 109(3), 2053-2065.

Abramo, G., D'Angelo, C.A., & Di Costa, F. (2010). Citations versus journal impact factor as proxy of quality: Could the latter ever be preferable? *Scientometrics*, 84(3), 821-833.

Adams, J. (2005). Early citation counts correlate with accumulated impact. *Scientometrics,* 63(3), 567-581.

Baumgartner, S., & Leydesdorff, L. (2014). Group-based trajectory modelling (GBTM) of citations in scholarly literature: Dynamic qualities of "transient" and "sticky" knowledge claims. *Journal of the American Society for Information Science and Technology*, 65(4), 797-811.

Bornmann, L., & Daniel, H.D. (2008). What do citation counts measure? A review of studies on citing behavior. *Journal of Documentation*, 64(1), 45–80.

Bornmann, L., Leydesdorff, L., & Wang, J. (2014). How to improve the prediction based on citation impact percentiles for years shortly after the publication date? *Journal of Informetrics*, 8(1), 175-180.

Bruns, S. B., & Stern, D. I. (2016). Research assessment using early citation information. *Scientometrics,* 108(2), 917-935.

Garfield, E. (1972). Citation analysis as a tool in journal evaluation. *Science,* 178, 471-479.

Glänzel, W., Schlemmer, B., & Thijs, B. (2003). Better late than never? on the chance to become highly cited only beyond the standard bibliometric time horizon. *Scientometrics, 58*(3), 571-586.
Levitt, J. M., & Thelwall, M. (2011). A combined bibliometric indicator to predict article impact. *Information Processing and Management, 47*(2), 300-308.
Li, X., Thelwall, M., & Giustini, D. (2012). Validating online reference managers for scholarly impact measurement. *Scientometrics*, 91(2), 461-471
Mingers, J. (2008). Exploring the dynamics of journal citations: modelling with S-curves. *Journal Operational Research Society*, 59 (8), 1013-1025.
Priem, J., Taraborelli, D., Groth, P., & Neylon, C. (2010). *Alt-Metrics: A Manifesto*. Retrieved from http://altmetrics.org/manifesto/ Last accessed 6 Nov. 2018.
Rousseau, R. (1988), Citation distribution of pure mathematics journals. In: *Egghe, L., Rousseau, R. (Ed.) Informetrics,* Belgium: Diepenbeek, *87/88*, 249-262, *Proceedings 1st International Conference on Bibliometrics and Theoretical Aspects of Information Retrieval*.
Shema, H., Bar-Ilan, J., & Thelwall, M. (2014). Do blog citations correlate with a higher number of future citations? Research blogs as a potential source for alternative metrics. *Journal of the Association for Information Science and Technology*, 65(5), 1018-1027.
Stegehuis, C., Litvak, N., & Waltman, L. (2015). Predicting the long-term citation impact of recent publications. *Journal of Informetrics, 9*(3), 642-657.
Stern, D.I. (2014). High-ranked social science journal articles can be identified from early citation information. *PLoS One*, 9(11), 1-11.
Stringer, M.J., Sales-Pardo, M., & Amaral, L.A.N. (2008). Effectiveness of journal ranking schemes as a tool for locating information. *PLoS ONE, 3*(2) doi:10.1371/journal.pone.0001683
Sud, P., & Thelwall, M. (2014). Evaluating altmetrics. *Scientometrics*, 98(2), 1131-1143.
Thelwall, M., & Sud, P. (2016). Mendeley readership counts: An investigation of temporal and disciplinary differences. *Journal of the Association for Information Science and Technology, 67*(12), 3036-3050.
Thelwall, M., Haustein, S., Larivière, V., & Sugimoto, C. R. (2013). Do altmetrics work? Twitter and ten other social web services. *PLoS ONE*, 8(5), e64841.
Zeileis, A., (2004). Econometric computing with HC and HAC covariance matrix estimators. *Journal of Statistical Software*, 11(10), 1-17.
Wang, D., Song, C., & Barabási, A. (2013). Quantifying long-term scientific impact. *Science, 342*(6154), 127-132.
Wang, J. (2013). Citation time window choice for research impact evaluation. *Scientometrics, 94*(3), 851-872.
Wang, J., Mei, Y., & Hicks, D. (2014). Comment on "quantifying long-term scientific impact". *Science, 345*(6193).